\documentclass[letter]{aa} 
\usepackage{graphicx}
\usepackage{txfonts}
\begin{document}
 \subtitle{Dust-continuum mapping of an infrared dark cloud with P-ArT\'eMiS} 
 \title{Evidence of triggered star formation in G327.3-0.6}
   
   \author{V. Minier
          \inst{1}\fnmsep\thanks{We thank the staff of the APEX telescope for their support.  
          APEX is a collaboration between the  Max Planck Institut f\"ur Radioastronomie, ESO and Onsala Space Observatory}
          \and
          Ph. Andr\'e \inst{1}     
          \and
          P. Bergman \inst{2, 3}
          \and
          F. Motte \inst{1}
          \and
          F. Wyrowski \inst{4}          
          \and
          J. Le Pennec \inst{1}
          \and
          L. Rodriguez \inst{1}
          \and
          O. Boulade  \inst{1}
          \and   
          E. Doumayrou \inst{1}
          \and
          D. Dubreuil \inst{1}
          \and
          P. Gallais \inst{1}
          \and
          G. Hamon \inst{1}
          \and
          P.-O. Lagage  \inst{1}
          \and
          M. Lortholary \inst{1}
          \and
          J. Martignac  \inst{1}
           \and
          V. Rev\'eret \inst{1, 3}
          \and
          H. Roussel  \inst{5}
          \and
          M. Talvard  \inst{1}
          \and
          G. Willmann  \inst{1}
          \and
          H. Olofsson \inst{2}
                    }
          
   \institute{Laboratoire AIM, CEA Saclay-Universit\'e Paris Diderot-CNRS, DSM/Irfu/Service d'Astrophysique, 91191 Gif-sur-Yvette, France \\
              \email{vincent.minier@cea.fr}
         \and
             Onsala Space Observatory, Chalmers University of Technology, 439 92 Onsala, Sweden
        \and
        European Southern Observatory, Casilla 19001, Santiago 19, Chile   
        \and
         Max Planck Institut f\"ur Radioastronomie, Auf dem H\"ugel 69, D-53121 Bonn, Germany
         \and
            Institut d'Astrophysique de Paris, UPMC (Universit\'e Paris 6), 98b Bd Arago, 75014 Paris, France            
             }


 
  \abstract
   {}
   {Expanding HII regions and propagating shocks are common in the environment of young high-mass star-forming complexes. They can compress a pre-existing molecular cloud  and trigger the formation of dense cores. We investigate whether these phenomena can explain the formation of high-mass protostars within an infrared dark cloud located at the position of G327.3-0.6 in the Galactic plane, in between two large infrared bubbles and two HII regions.}
   {The region of G327.3-0.6 was imaged at 450 $\mu $m with the CEA P-ArT\'eMiS bolometer array on the Atacama Pathfinder EXperiment telescope in Chile. APEX/LABOCA and APEX-2A, and Spitzer/IRAC and MIPS archives data were used in this study.}
   {Ten massive cores were detected in the P-ArT\'eMiS image, embedded within the infrared dark cloud seen in absorption at both 8 and 24 $\mu$m. Their luminosities and masses indicate that they form high-mass stars. The kinematical study of the region suggests that the infrared bubbles expand toward the infrared dark cloud. }
   {Under the influence of expanding bubbles, star formation occurs in the infrared dark areas at the border of HII regions and infrared bubbles.}


   \maketitle
%

\section{Introduction}

A necessary prerequisite to star formation is the existence of dense cores of self-gravitating gas. How the dense cores form, evolve and give birth to new stars remains a matter of debate for high-mass star formation (e.g., Motte et al. 2007).  Possible phenomena responsible for their formation include external events such as expanding HII regions and propagating shocks, which are common in the environment of young high-mass star-forming complexes (e.g., Zavagno et al. 2007) . The ionised front of an HII region can compress a pre-existing molecular cloud and trigger the formation of dense cores  (e.g., Purcell et al. 2009). Stellar winds or shocks from supernovae can also lead to the formation of similar objects (e.g., Koo et al. 2008). The Galactic Legacy Infrared Mid-Plane Survey Extraordinaire (GLIMPSE; Churchwell et al. 2006) detected many infrared bubbles in the Galactic plane close to active sites of high-mass star formation. The infrared emission tracing the bubbles was interpreted as a signature of PAH emission in the photo-dissociation regions (PDRs) at the borders of HII regions. The bubbles may also be inflated by strong stellar winds. 

We present the first  450 $\mu$m map of massive protostar progenitors within the high-mass star-forming region near the hot molecular core G327.3-0.6 (Wyrowski et al. 2006) at the projected extremity of an infrared dark cloud (Fig. 1). At a distance of 2.9 kpc (Bergman 1992), G327.3-0.6 is surrounded by two large scale infrared bubbles and two HII regions, as shown in Fig. 1. The objective of this letter is to {\bf{(1)}} provide a census of the objects in an early phase of star formation that are embedded in the infrared dark areas and {\bf{(2)}} investigate the relationship between early phases of high-mass star formation and the influence of the expansion of nearby infrared bubbles and HII regions.

   \begin{figure*}
   \centering
   \includegraphics[scale=0.5]{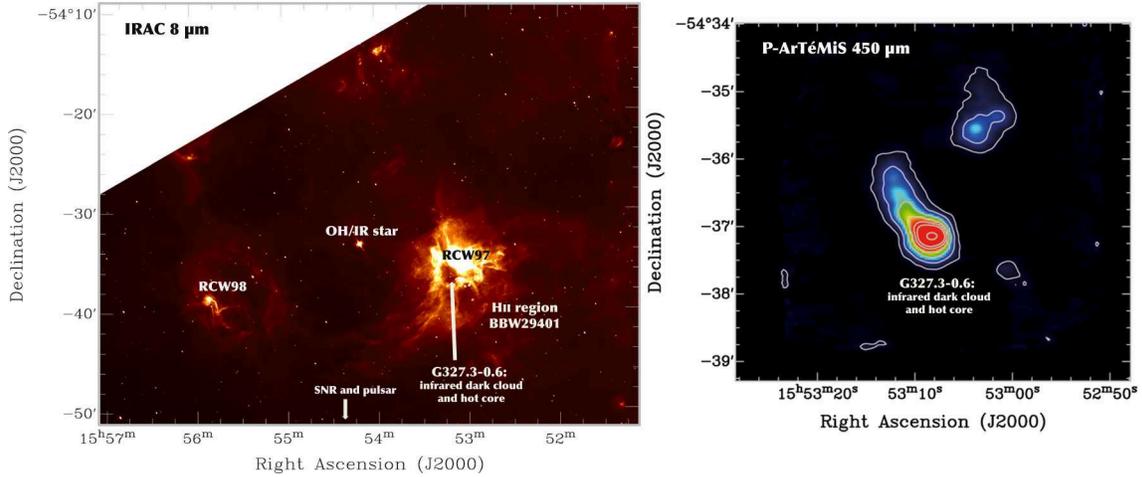}
   \caption{Overview of G327. Left: Spitzer/IRAC image at 8 ${\mu}$m of the large scale infrared bubble with G327.3-0.6 and RCW97 at the border. Right: P-ArT\'eMiS 450-${\mu}$m emission. The contours are the 450-${\mu}$m emission at levels of 4, 10, 20, 40, 50, and 90\% of the 140 Jy beam$^{-1}$ peak flux. The rms is about 0.5 Jy beam$^{-1}$ on the map.}
              \label{overview}%
    \end{figure*}

\section{P-ArT\'eMiS 450 $\mu$m continuum imaging}

\subsection{Observations} 

The region of G327.3-0.6 was imaged at 450 $\mu $m in November 2007 with the P-ArT\'eMiS bolometer array on the Atacama Pathfinder Experiment (APEX) telescope located at an altitude of 5100 m at Llano de Chajnantor in Chile. The P-ArT\'eMiS detector is a $16\times16$-pixel prototype of the ArT\'eMiS large-format bolometer camera being built by CEA Saclay for APEX (e.g., Talvard et al. 2008). 
Three individual maps, corresponding to a total effective integration time of 1.4 hours, were obtained with P-ArT\'eMiS at $450~\mu$m toward the G327.3-0.6 region using a total-power, on-the-fly scanning mode. Each of these maps consisted of a series of scans in either Azimuth or Right Ascension taken with a scanning speed of 16$^{\prime \prime }$ s$^{-1}$. The cross-scan step between consecutive scans was 2$''$. 

The atmospheric opacity at zenith was monitored by taking skydips with P-ArT\'eMiS and was found to be between 0.6 and 0.8 at $\lambda = 450~\mu$m, corresponding to precipitable water vapour amount between 0.5 mm and 0.7 mm. No dedicated pointing model was derived for P-ArT\'eMiS, but the pointing corrections proved to be very similar i.e., to within $\sim$10$''$ of the corrections appropriate for the APEX-2A instrument. Pointing, focus, and calibration measurements were achieved by taking both short ``spiral'' scans and longer on-the-fly beam maps of Mars and Saturn. We estimated the absolute pointing accuracy to be $\sim$5$''$. The flux amplitude was to within $\sim30\%$ uncertainty. As estimated from Mars maps, whose angular diameter was $\sim$14$''$, the main beam had a full width at half maximum of $\sim 10''$ and contained $\sim50\%$ of the power. Offline data reduction, including baseline subtraction, removal of correlated skynoise and 1/f noise, and subtraction of uncorrelated 1/f noise using a method exploiting the high level of redundancy in the data, were performed with in-house IDL routines (Andr\'e et al. 2008). 

\subsection{Results and analysis} 

The 450 $\mu$m emission originates in a large infrared dark cloud ($2\times1$ pc$^2$) near G327.3-0.6 and at the south-eastern border of the RCW97 HII region (Fig. 1). Spitzer IRAC and MIPS images as well as LABOCA/APEX and APEX-2A archives observations were used in addition to the P-ArT\'eMiS data set. 

Ten submillimetre sources (SMM) were identified using a version of the Gauss-clumps program and the MRE multi-resolution program based on wavelet transforms (Starck \& Murtagh 2006), which has been customised for continuum images (see Appendix B in Motte et al. 2007). This method allowed us to first detect and then characterise SMM2, 4, 7, and 8, which were not identified by eye in the original 450 $\mu$m map (Fig. 1). We used MRE to complete a series of filtering operations and provided ÒviewsÓ of the image on different spatial scales. A cutoff angular size of $24''$ (i.e., 0.3 pc at 2.9 kpc), on which scale SMM2, 4, 7, and 8 were resolved, was estimated. All wavelet planes of the original image up to this scale were summed to create images such as those shown in Fig. 2. We note that the original image can be described by the sum of all wavelet views plus the smoothed image (last plane containing all remaining scales). Table 1 presents the SMM source characteristics ($F_{peak}$, the peak flux in the original map; $F_{24''}$, the peak flux in the filtered map; $FWHM$, the full width half maximum size; $F_{int}$, the integrated flux) from both the filtered map (Fig.  2) and the original map (Fig. 1). The brightest source, SMM1, coincides with the position of the hot core and an infrared source (IRS1 in Fig. 2). SMM4 coincides with a 8 and 24 $\mu $m infrared source (IRS2 in Fig. 2). Other sources do not have any infrared source counterpart in the MIPS or IRAC image. 

   \begin{figure}
   \centering
   \includegraphics[scale=0.47, angle=0]{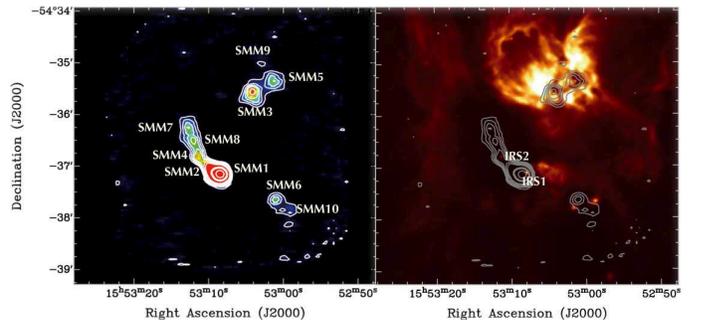}
   \caption{Left: The image and contours  represents the filtered P-ArT\'eMiS 450-${\mu}$m emission  where the emission over an area larger than $24''$ has been filtered. The contours are the 450-${\mu}$m emission at levels of 2, 4, 6, 8, 12, 16, 40, and 75\% of the 101 Jy beam$^{-1}$ peak flux. The rms is about 0.4 Jy beam$^{-1}$ on the map. Right: Zoomed view of  the IRAC 8-${\mu}$m emission image. Contours represent the filtered P-ArT\'eMiS  450-${\mu}$m emission.}
              \label{overview}%
    \end{figure}

Assuming optically thin dust continuum emission at 450 $\mu $m, the gas masses $M$ of the sources were derived from the measured flux densities $F_{450~\mu {\rm m}}$ (either $F_{int}$ or $F_{24''}$) using
\begin{equation}
M = \frac{F_{450~\mu {\rm m}} D^2} {\kappa_{450~\mu {\rm m}}~ B_{450~\mu{\rm m}}(T_{\rm d})}, 
\end{equation}
where $D$ is the distance to the source, $\kappa_{450~\mu {\rm m}}$ is the dust opacity per unit (gas + dust) mass column density at $\lambda = 450~\mu {\rm m}$, and $B_{450~\mu {\rm m}}(T_{\rm d}) $ is the Planck function $B_\nu (T_{\rm d}) $ for the dust temperature $T_{\rm d}$. We adopted $\kappa_{450~\mu {\rm m}} = 0.04~\rm {cm}^{2}~\rm {g}^{-1}$, which is consistent with the dust opacity law estimated by Hill et al. (2007) and appropriate in regions of moderately high gas densities ( $n_{{\rm H}_2} \sim10^5~\rm {cm}^{-3}$). The mean dust temperatures of SMM1, 2, 3, 5, 6, 9, 10 were estimated from the observed 450 and 870 $\mu $m flux density ratio by assuming optically thin dust emission with an emissivity index $\beta = 2$. The 870 $\mu $m flux density was derived from LABOCA/APEX observations. For SMM1, the hot core, and SMM 3, 5, and 9 in the RCW97 PDR, the temperature estimates are probably lower limits, corresponding to upper mass limits. For three sources, SMM4, 7, and 8, the flux density ratio map did not provide any accurate measurement. A mean temperature for this part of the dark cloud was estimated to be $\sim20$ K and used for these three sources (see Table 1). 

\begin{table*}
\begin{minipage}[]{\columnwidth}
\caption{Dense cores detected in the G327 region.}             
\label{table:1}      
\centering                          
\renewcommand{\footnoterule}{}
\begin{tabular}{c c c c c c c c c c c}        
\hline\hline                 
Fragment & RA  & Dec  & $F_{peak}$ \footnote{using the original map in Fig. 1.} & $F_{24''}$\footnote{using the filtered map in Fig. 2. SMM2 needs confirmation.} & \it FWHM & $F_{int}$ & $T_d$\footnote{$T_d$ value is uncertain within 5 K.} & $M_{int}^{FWHM}$\footnote{with a factor 2 of uncertainty on either side of the mass value.} & $M_{env}^{6000 AU}$$^{b,d}$\\
name &  J2000 & J2000 & [Jy/beam] &  [Jy/beam] & [\arcsec~$\times$~\arcsec] & [Jy] & [K]  &  [M$_{\odot}$] &  [M$_{\odot}$] \\
\hline      
SMM1 & 15:53:08.6 & -54:37:09 & 140 & 101 & $19 \times 16$ & 440 & 20  & 3769 & 184 \\
SMM2$^b$ & 15:53:10.1 & -54:37:01 & - & 19 & $16 \times 11$ & 35$^b$ & 25 & 197$^b$ & 23 \\
SMM3 & 15:53:04.0 & -54:35:34 & 24 & 13 & $25 \times 22$ & 132 & 15 & 2129 & 45 \\
SMM4$^b$ & 15:53:11.5 & -54:36:46 & - & 11 & $20 \times 11$ & 25$^b$ & 20 &  214$^b$ & 20 \\ 
SMM5 & 15:53:01.4 & -54:35:20 & 17 & 8 & $28 \times 21$ & 100 &  40 &  265 & 4  \\
SMM6 & 15:53:00.9 & -54:37:40 & 10 & 8 & $15 \times 14$ & 22 & 15 &  354 & 27  \\
SMM7$^b$ & 15:53:12.8 & -54:36:11 & - & 5 & $14 \times 13$ & 9$^b$ & 20  & 77$^b$ & 9 \\ 
SMM8$^b$ & 15:53:12.1 & -54:36:31 & - & 3   & $11 \times 10$ & 3$^b$ & 20 &  25$^b$ & 5  \\ 
SMM9 & 15:53:03.3 & -54:34:58 & 8 & 3 & $34 \times 21$ & 60 &  20 & 514 & 5  \\
SMM10 & 15:52:59.1 & -54:37:52 & 7 & 2 & $19 \times 16$ & 20 & 20  & 171 & 4 \\
\hline                                   
\end{tabular}   
\end{minipage}
\end{table*}

Mass estimates of the SMM sources were derived from both the peak fluxes and the integrated fluxes over their deconvolved sizes.  There was an uncertainty of a factor of 2 on either side, mainly because of uncertain value of $\kappa_{450~\mu {\rm m}}$. SMM sources have masses that vary from 25 to 3800 M$_{\odot}$ (Table 1). A second set of mass values, M$_{env}^{6000 AU}$, was also derived, which correspond to the estimate of the gas envelope masses of possibly dominant protostellar objects embedded within each SMM source. The diameter of the protostellar envelope is set to be 6000 AU (the typical fragmentation length-scale observed in high-mass star-forming regions - e.g., Longmore et al. 2006). The value of  M$_{env}^{6000 AU}$ was obtained by first calculating the mass corresponding to the peak flux density in the 10$^{''}$ beam and then applying a scaling factor to that mass, assuming a $\rho{\propto}r^{-2}$ density distribution so that the measured mass scales as the aperture size. The value of M$_{env}^{6000 AU}$ varies from 4 to 190 M$_{\odot}$ (see Table 1). The overall mass of the dark cloud was estimated to be $\sim1.7\times10^5$ M$_{\odot}$ assuming a mean dust temperature of 20 K and taking into account the extended emission. 

SMM1 and SMM4 are associated with infrared sources at both 8 and 24 $\mu$m (Fig. 2). Based on fits to approximate their spectral energy distribution from 24 to 870 $\mu$m, the luminosities of SMM1 and SMM4 are estimated to be within the ranges of  $5{\times}10^4-10^5$ L$_{\odot}$ and $3{\times}10^3-10^4$ L$_{\odot}$, respectively. The luminosities of the other SMM sources were not calculated because of  the lack of mid-infrared emission at 24 $\mu{\rm m}$ in the Spitzer/MIPS image. However, their luminosities were estimated to be within $3{\times}10^2-10^4$ L$_{\odot}$ using their flux at 450 $\mu$m to extrapolate the luminosity values from those estimated for SMM1 and SMM4. The positions of SMM1 and SMM4 in an $M_{env}-L_{bol}$ diagram suggests that they are border-line Class 0/Class I objects that will evolve into massive stars of masses $M_* >~ 50$ M$_{\odot}$ and  $M_* >~ 15$ M$_{\odot}$, respectively (Fig. 3). The other SMM sources, which have no detected IR emission, might represent earlier phases such as highly embedded class 0 protostars or massive prestellar cores. Their mass estimates suggest that they contain enough material to form new high-mass stars. In particular, SMM3 might represent a candidate massive prestellar object that has formed at the border of the RCW97 HII region.       
      
\begin{figure}
   \centering
   \includegraphics[scale=0.4 , angle=0]{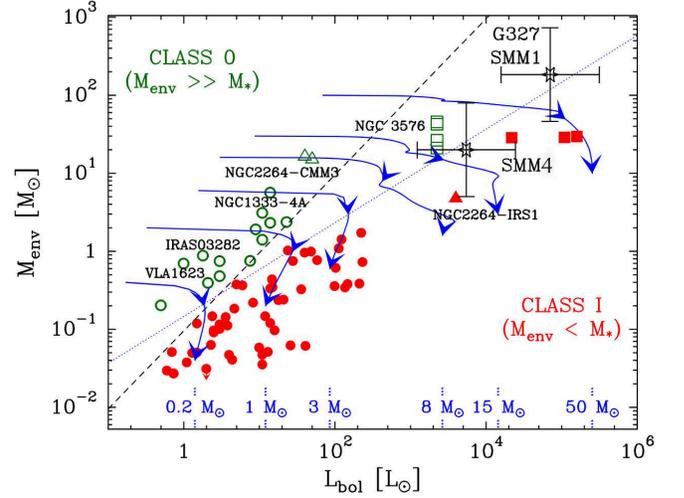}
   \caption{Envelope mass versus bolometric luminosity diagram comparing the locations of the protostellar
sources of G327.6-0.3 (stars with error bars) with the positions of low-mass Class I (filled circles), Class 0 objects (open circles), and high-mass protostars in NGC3576 (squares - Andr\'e et al. 2008), as well as intermediate-mass protostars in NGC 2264 (triangles - Maury et al. 2009). Model protostellar evolutionary tracks, computed for various final stellar masses assuming the accretion/luminosity history described in Andr\'e et al. (2008), are superimposed. Evolution proceeds from the upper left to the lower right as indicated by two arrows on
each track, plotted when 50\% and 90\% of the final stellar mass has been accreted, respectively. The straight lines show two $M_{env}-L_{bol}$ relations marking the 
conceptual border zone between the Class 0 ($M_{env} > M_{*}/{\epsilon}$) and the Class I ($M_{env} < M_{*}/{\epsilon}$) stage (where $\epsilon$ is the local star formation
efficiency). The dashed line is such that $M_{env}  \propto L_{bol}$, while the dotted relation follows $M_{env}  \propto L_{bol}^{0.6}$ as suggested by the accretion 
scenario adopted in the tracks.}
\end{figure}
      
\section{Triggered star formation in the dark cloud ?}

\begin{figure*}
   \centering
   \includegraphics[scale=0.9, angle=0]{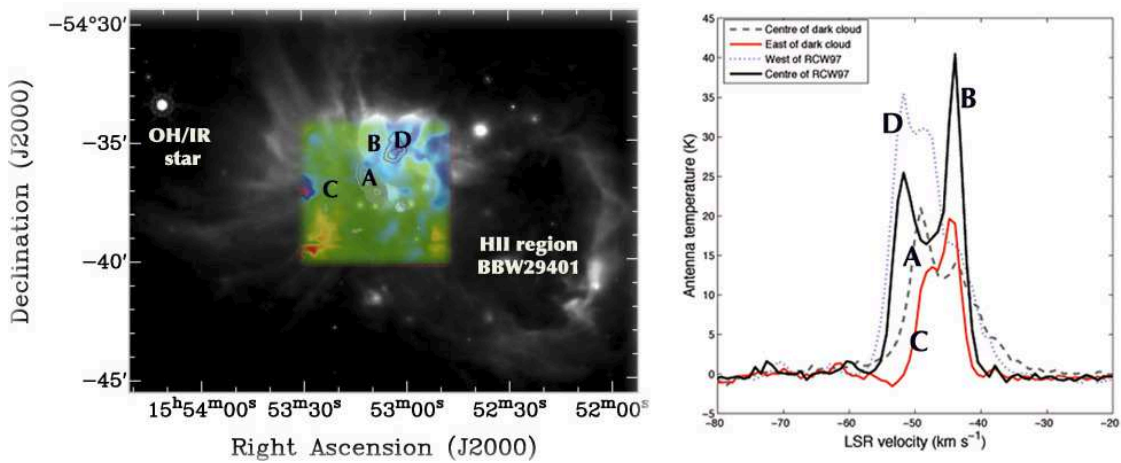}
   \caption{Left: First moment map of the CO(3-2) line emission (colour image) superimposed on the MIPS 24-$\mu$m emission image (black and white image). Blue to red colours correspond to blue- and red-shifted velocities. The  grey contours represent the P-ArT\'eMiS 450-${\mu}$m emission.  Right: Spectra taken at 4 positions in the first moment map: A corresponding to SMM4 in the infrared dark cloud, B to RCW97, C to the border of the infrared bubble, and D to west of RCW97 between SMM3, 5 and 9.}
   \end{figure*}

The MIPS 24 $\mu$m emission image clearly exhibits two infrared bubbles (Fig. 4). 
The eastern one corresponds to the large bubble seen in the IRAC image (Fig. 1). It is centred on a OH/IR type star (IRAS15502-5424) that is associated with circumstellar outflows of blueshifted OH maser velocities of between -20 and -47 km~s$^{-1}$ (Te Lintel Hekkert et al. 1991). The star LSR velocity is estimated to be -8 km~s$^{-1}$. Interestingly, many molecular clouds and high-mass star-forming regions (SFO75; RCW98) are observed around the bubble with LSR velocities close to the systemic velocity of G327.3-0.6 (Russeil \& Castets 2004; Urquhart et al. 2006). Two HII regions are also present near G327.3-0.6. To the north west of the infrared dark cloud is RCW97. To the south west, lies the HII region known as BBW29401, which is the brightest object in the H$\alpha$ image at -48 km~s$^{-1}$ (Russeil 2003). The second largest 24 $\mu$m infrared bubble encompasses BBW29401 (Fig. 4).  

The region of G327.3-0.6 was imaged in $^{12}$CO$(3-2)$ using the APEX-2A receiver on the APEX telescope (Wyrowski et al. 2006) . Using these archives observations, a first moment map and line spectra were created to determine whether the infrared bubbles expand, compress the infrared dark cloud, and trigger star formation in G327.3-0.6. The velocity field around G327.3-0.6 can be divided, to a first approximation, across an east-west axis in two velocity colours: a blueshifted velocity ($\sim-50$ km~s$^{-1}$) domain to the west, mainly located west of RCW97 (position D in Fig. 4); and a redshifted velocity ($\sim-46$ km~s$^{-1}$) domain to the east (position C in Fig. 4). Along the eastern border of the infrared dark cloud traced by the curved infrared filament, the velocity remains at the value of -46 km~s$^{-1}$. The CO line toward the RCW97 HII region exhibits a double profile with peaks at -50 and -46 km~s$^{-1}$ (position B in Fig. 4). Moving to the west and south-west of RCW97, the velocity shifts toward -50 km~s$^{-1}$. The morphology of RCW97 in the infrared also divides into two regions, with SMM3, 5, and 9 in between (Fig. 2). The CO spectrum at the position of SMM4 exhibits multiple velocity components of velocity between -50 and -40 km~s$^{-1}$ (position A in Fig. 3). The 4~km~s$^{-1}$ range of velocity peaks might indicate that the bubbles expand if they are not located at the distance of G327.3-0.6, but in its foreground and background. Similar profiles of CO spectra are modelled by Hennebelle et al. (2006) to explain a rapid increase  in external pressure driving a compression wave in the Coalsack region, but of a far smaller velocity range. In any case, the CO line emission is probably self-absorbed by the cloud, which to some extent could also explain the spectra.  

The kinematical properties and the detection of the infrared bubbles at both 8 $\mu$m, a signature of PAH, and at 24 $\mu$m, a signature of heated interstellar dust, suggest that they possibly trace the interface between expanding fronts traced by PAH emission and the molecular cloud traced by thermal dust emission and gas.

\section{Conclusions}

The P-ArT\'eMiS image has detected massive cores embedded within the infrared dark cloud seen in absorption at both 8 and 24 $\mu$m. Three SMM sources are detected at the western border of the RCW97 HII region. These SMM sources also coincide with infrared dark areas. The luminosity estimates of SMM sources indicate that the massive cores host luminous objects and therefore form high-mass stars. 

The RCW97 HII region and the infrared dark cloud emit CO lines within the same velocity range. A possible interpretation is that both the infrared dark cloud and RCW97 are embedded in the same parent molecular cloud. Under the influence of expanding bubbles, star formation now occurs on the edge of RCW97, in the infrared dark cloud and at the border of HII region itself. In conclusion, circumstantial evidence of triggered star formation has been observed toward G327.3-0.6. 

\begin{acknowledgements}
This research work has been funded by the Agence Nationale de la Recherche (ANR project: ArT\'eMiS). Part of this work is based  on observations made with the Spitzer Space Telescope, which is operated byJPL and Caltech under a contract with NASA.
\end{acknowledgements}

\end{document}